\documentclass[]{aa}
\usepackage{graphicx}
\usepackage{natbib}
\bibpunct{(}{)}{;}{a}{}{,}
\usepackage{txfonts}
\begin{document}
\title{Indirect evidence for short period magnetic cycles in W UMa stars}
\subtitle{Period analysis of five overcontact systems}

   \author{T. Borkovits\inst{1}, M.M. Elkhateeb\inst{2}, Sz. Csizmadia\inst{3}, J. Nuspl\inst{3},
            I.B. B\'\i r\'o\inst{1}, T. Heged\"us\inst{1}, R. Csorv\'asi\inst{4,5}}

   \offprints{T. Borkovits, \newline \email borko@alcyone.bajaobs.hu}

   \institute{Baja Astronomical Observatory of B\'acs-Kiskun County,
             H--6500 Baja, Szegedi \'ut, Kt. 766, Hungary
	     \and
	     Dept. of Astronomy, National Research Institute of Astronomy and Geophysics
             (NRIAG), Helwan, Cairo, Egypt
             \and
   	     Konkoly Observatory of the Hungarian Academy of Sciences,
             H--1525 Budapest, Pf. 67, Hungary
	     \and
             Department of Optics \& Quantum Electronics \& Astronomical Observatory,
	     University of Szeged, H--6701 Szeged, Pf. 406, Hungary
	     \and
	     Guest Observer at Piszk\'estet\H o Station, Konkoly Observatory, Hungary}

   \titlerunning{Short period magnetic cycles in W UMa stars}
   \authorrunning{Borkovits et al.}

   \date{Received  / Accepted }

   \abstract{Complex period variations of five W~UMa type binaries (\object{AB~And},
              \object{OO~Aql}, \object{DK~Cyg}, \object{V566~Oph}, \object{U~Peg}) were investigated
              by analyzing their O--C diagrams, and several common features were found. Four of
              the five systems show secular period variations at a constant rate on the order of
              $|\dot{P}_\mathrm{sec}/P|\sim10^{-7}$~y$^{-1}$. In the case of \object{AB~And}, \object{OO~Aql},
              and \object{U~Peg} a high-amplitude, nearly one-century long quasi-sinusoidal pattern was also
              found. It might be explained as light-time effect, or by some magnetic phenomena, although
              the mathematical, and consequently the physical, parameters of these fits are very problematic,
              as the obtained periods are very close to the length of the total data range. The most interesting
              feature of the studied O--C diagrams is a low amplitude ($\sim2-4\times10^{-3}$ d)
              modulation with a period around 18--20 yr in four of the five cases.
              This phenomenon might be indirect evidence of some magnetic cycle in late-type 
              overcontact binaries as an analog to the observed activity cycles in RS~CVn systems.


}

\maketitle

\section{Introduction}

The moments of minima are one of the most fundamental observables of eclipsing binary systems and,
due to their relatively easy measurement, a huge pile of data of this type was collected by observers
during the last century. However, so far there is no a commonly accepted and straightforward
interpretation of the O--C diagrams constructed from them. Additional difficulties arise when
one tries to disentangle and identify the physical mechanisms
causing the observed period variations although this would be the aim.
For example, in spite of the fact that the period variations are common features of contact
binary stars \citep{kreiner77,maceronivantveer96}, their origin is not well understood yet.

These observed variations can be classified into two subsets:
(i) in the first part are the apparent period variations due to light-time effect (LITE)
caused by a distant third or further body\footnote{ The other fundamental
class of the apparent period variations, namely, the effect of the apsidal motion (AM)
is not considered here due to the circular orbit of the systems.}, and
(ii) the second part comprises the inherent physical period variations.
The latter may also be of various types, e.~g. caused by the evolution of the system, mechanical
and/or thermodynamical effects, or due to variable magnetic activity. Of course, a mixture
of the different sources can be active at the same time, generating a very complex variation
in the orbital period.

In respect to the involved time scales we can distinguish the long-term variations
(on a nuclear or thermal time scale) from the short-term variations characterised
by a decade-long cycle length. In our recent study we focus on the short--term
orbital period variations of several W~UMa systems, because the real O--C observations can reveal
only these effects and the long--term ones require a different (statistical) approach.

\subsection*{Long-term effects}

Recently, the internal structure and evolution of contact W~UMa systems are not well--known
or else are poorly understood. The \emph{evolution theories} of these binary stars yield secular as well as
shorter time scale quasi-cyclical orbital period variations. Although these theories suffer
from a lot of problems their predictions can be used as guidelines
\citep{kreineretal03,kahler04} to better understand the behaviour of these systems.

All the currently available theories predict a secular
orbital period increase of such systems caused by mass transfer between the
components. Some theories, e.~g. the Thermal Relaxation Oscillation (TRO) \citep{lucy76,webbink76} or
discontinuity (DSC) theory \citep{lubowshu77},
give rise to different periodic and secular variations of the orbital period and of the
observable light curves. However, \citet{mochnacki81,kahler86}, as well as \citet{hazlehurst01}
showed that none of these theories agree with the observations or seem to violate the second
law of thermodynamics.
The Angular Momentum Loss (AML) theories are based on the assumption of the angular
momentum loss via magnetic breaking on stellar winds \citep{vantveer79} or on
the magnetic fields of the companion star \citep{mochnacki81}, or even on the simple
mass and angular momentum loss via stellar wind \citep{vantveermaceroni89}.

\citet{kahler04} integrated the evolutionary equation of a model contact system and found that the systems
are oscillating both with and without losing their contact configuration. During
these oscillations the period shows (quasi-)periodic variations on a time
scale of several million years. The origin of this period variation is once again
mass transfer between the components.
Note that \citet{qian01a} suggested that systems with $q<0.4$
show secular period decrease and systems with $q>0.4$ secular period
increase ($q$ is the mass ratio), an effect that - if real - has not yet been
interpreted theoretically.

Independent of any uncertainties in them, all of these theories predict the
period variations on a time-scale comparable to the nuclear evolution time-scale of the
primary components i.~e. some billion years which is far from what is observed in the O--C
diagrams.

\subsection*{Short-term effects}

Short-term effects are the most exciting ones for us because they can be studied through the observed times of
minima and the O--C diagrams.

The \emph{LITE} that cause apparent orbital period variations was studied
in many contact binary systems, and one of the most recent extended studies
was published in \citet{borkohege96}, also includes 18 Algol and contact
systems. They found that only four of the 18 cases studied might have a third
(or fourth) companion with high probability. Remarkably two of the studied
five W~UMa systems, namely \object{AB And} and \object{AK Her}, could have a third component compared to the
two Algol type systems from the fifteen ones considered.

However, \citet{hendrymochnacki98} found that \object{AK Her} does
not show spectroscopic evidence for a third companion in the system.
Note that this star has a wider, visual companion at a separation of
$\rho=4\farcs7$, nevertheless this companion cannot be responsible for the
decades-scale cyclic variation of the O--C. Since the time coverage of
their observations was only a few days, the gamma velocity was not
changed, while the lack of the third body's spectroscopic line can be
explained by a low luminosity object (e. g. a white dwarf).  They also
found from spectroscopy that \object{SW Lac} and \object{V502 Oph} have a
third body, although the O--C diagrams did not show their presence clearly. 
These facts might be explained by a high inclination
effect. This contradiction between the O--C diagrams and the
spectroscopic observations has not been solved yet.

Note that the light-time effect explanations of the O--C diagrams are
uncertain in those cases when less than one revolution of the third body
could be observed till now. One important example for this latter
statement can be the case of \object{AK Herculis}
\citep{rovithislivaniouetal99}.  At the same time, in the case of
\object{VW Cephei} the period of the LITE orbit deduced from the O--C
diagram is in very good agreement with the astrometric observations
\citep{hershey75,heintz93}, although the amplitude of the LITE and,
consequently, the projected semi-major axis of the third body orbit was
found to be inconsistent with these other measurements
\citep[see][]{kaszasetal98}. Nevertheless, independent of the
possible reasons for this inconsistency, it seems evident that at
least a significant part of the cyclic O--C variation arises from the
astrometrically, and interferometrically identified tertiary. Although 
it is clear that the LITE studies should be improved, it
is not an easy task considering that the typical time-scale of the
revolution period of the third body is as long as several decades or
centuries. Thus, it should be emphasised that independent spectroscopic
affirmations are inevitable before one can approve any LITE solution constructed
from O--C diagrams.

 A further reason for period variations on a short time-scale can be the
 \emph{magnetic activity} of the contact binary system. There are significant
pieces of evidence for their magnetic activity, such as
\begin{enumerate}
\item[(i)] Doppler-tomography \citep{barnesetal04,hendrymochnacki00,maceronietal94};
\item[(ii)] O'Connell-effect and their spot-explanation in many systems;
\item[(iii)] their X-ray emission deviations \citep{stepienetal01};
\item[(iv)] $H_\alpha$ emission in some W UMa systems \citep[e. g.][]{hrivnaketal95,csizmadiaetal04}.
\end{enumerate}
From an extended study of the period variations of a sample containing roughly 100 Algol
systems, \citet{hall98} concluded that the spectral type of systems showing cyclic period variations
is later than F5. This earlier hypothesis was supported during the last decade
by several investigations.

Magnetic activity is a manifestation of strong magnetic fields present in the system, which
interact with the matter of stars in plasma state and could change the
inertia tensor of component(s) through this interaction. Since the inertia tensor is a part
of the total angular momentum required to be constant, the variation in the magnetic field
structure might influence the orbital period of the system in this way.
This effect is known as ``the interchange of the magnetic and kinetic
energy'' \citep{applegate92,lanzarodono02}. The time-scale
of this so-called Applegate-mechanism is several years or decades, and its
amplitude may be as large as $0.006$ days.
This value is about 10-20 times higher than the accuracy of {\it one} minimum time observation.

Other reasons for period variations can also be imagined, e. g. sudden mass exchange via large
flares, gaining a planet, etc. Time-dependent variable long-scale current in
the common convective envelope could also cause variable orbital periods, while
all of these effects can affect the O--C diagram.

Since now more minima observations are available than a decade ago and further theoretical
calculations on the perturbational effects of a third body were carried out
\citep[e.~g.][]{borkoetal03}, we decided to improve and re-analyse the
O--C diagrams of some eclipsing contact binary stars. In Sect.~2 we describe
the method and in Sect.~3 we study these systems individually. The variations found in the
O--C diagrams are discussed in Sect.~4 and our conclusions can be found in Sect.~5.

\section{General considerations on the analysis of the O--C diagrams \label{Sec:O-Cgen}}
Finding a common explanation of the possible complex behaviour of O--C diagrams
is not an easy task as it requires careful detailed analysis and  modelling of the data set.
It is well-known that the quite general model of O--C data as the modulation of a secular (parabolic) trend
with some periodic additional variations due to different physical effects can usually describe most of the observed
O--C diagrams. In this representation, the secular part is connected to long--term evolution effects (mainly to mass exchange between
the components) and the periodic modulations can be connected to
e.~g. LITE caused by a distant third companion or, if no reasonable
third-orbit solution can be given, then some quasi-periodic magnetic effects or repeated mass--exchange processes.

Nevertheless, there are only a few systems where the presence of
a third companion has already been proved by independent measurements. In such systems
(e.~g. \object{VW~Cephei}, \object{XY~Leonis}), the tertiary components have a period several
times shorter than the total observational interval of the eclipsing minima range, and
these are clear cases of straightforward interpretation. Difficulty arises if the period of the cyclical term
(if any) is close to or longer than the observational window, especially
if any secular behaviour is also present. It is evident that an independent fitting
of the secular parabolic term and sinusoidal terms - as was done
in several earlier works - may give false results, and even a simultaneous
fitting does not necessarily improve the situation. Instead of the detailed description of
these problems, which are discussed and published elsewhere, we here give only two illustrations
to these statements. We show that such a simultaneous fit in the case what is of already one simple abrupt period change
can produce a reasonable -- secular plus periodic -- orbital period change solution.

In order to demonstrate this, we constructed different artificial times of minima data series.
All of them have the following ephemeris: $MIN_I=2\,445\,000.0+0.4556788E$ for $E<0$, and
$MIN_I=2\,445\,000.0+0.4556790E$ for $E>0$. The data were generated with a random accuracy of
about 0.0008 days and with random epochs (i.~e. random $E$ value). More accurate points than
0.0002 days were weighted as photoelectric data. Here we present two data series, one
which was calculated on the symmetric interval $-80\,000<E<80\,000$, while the second one
on the asymmetric one of $-110\,000<E<50\,000$.

The data series were analysed in a similar manner to the one described in \citet{borkoetal02}, 
i.~e. after calculating the O--C curve with a preliminary linear ephemeris, the final representation of the
O--C was searched for by a weighted linear least-squares fit in the form:
\begin{equation}
f=c_0+c_1E+c_2E^2+\sum_{i,j}a_{ij}\sin{j\nu_iE}+b_{ij}\cos{j\nu_iE},
\label{eq:O-Cgeneral}
\end{equation}
where $0\leq i\leq5$, $1\leq j\leq4$. The frequencies $\nu_i$ were kept fixed
during the individual LSQ runs, but an interval of frequencies was scanned
 and the best parameter set was selected according to the smallest $\chi^2$ test.

\begin{figure}
   \centering
  \resizebox{\hsize}{!}{\includegraphics{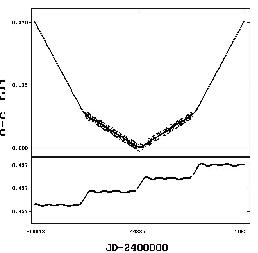}\includegraphics{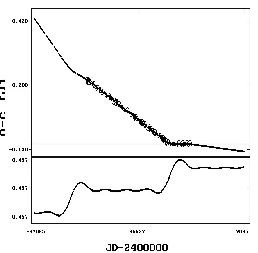}}
   \caption{The artificial O--C curves with simultaneously fitted least-square
parabolas and Fourier-curves (above) and the period curves (bottom). The  plotted time-intervals
are twice longer than the lengths of the data series.}
\label{fig:O-Cart}
    \end{figure}

In the present cases $i=1$, $j=1..3$ were applied, e.g. one fundamental frequency; and
its first two harmonics were used to obtain the very good fits, which are presented in
Fig.~\ref{fig:O-Cart}.
Furthermore, what is perhaps more suprising is that the ratio of the Fourier-coefficients,
which belong to different harmonics, almost exactly satisfy the condition of a light-time
orbit. Strictly speaking, we could find reasonable third body solutions in both cases, which
were almost independent of whether they were calculated only from
the $a_{1,2}$, $b_{1,2}$ coefficients \citep[][Chap.~V]{kopal78}, or the coefficients of
the second harmonics \citep{borkohege96} were also included.

Of course, the false results can be understood considering that removing the parabolic trend from the
data set the residual variations form a sawtooth-like periodic function, which can be represented
quite well by some low order terms of its Fourier--series. On the other hand, the solutions were significantly different for the two curves,
which sheds light on another problem of the simultaneous fitting. Namely, if only
approximately one period is covered by the observations, the coefficient obtained for
the quadratic term and the period of the periodic variation are strongly coupled, and
also depend on the length of the data.

In what follows we analyse the O--C curves of several contact binaries, some of
which clearly demonstrate this latter statement. However, we have to emphasise that in our cases these uncertainties
are connected  only to the longer period variations but not to the shorter ones which are the main targets
of our discussion here.

\section{Analysis of real systems}
We concentrated mainly on systems whose observation covers more than a half, or even a complete
century. The most important parameters of the analysed
five systems are listed in Table~\ref{table:Starmainparams}. The minima list of the
five stars are not included in this paper due to their enormous size. Nevertheless, the total
collection can be requested from the authors.

In several cases earlier observations are photographic or, more frequently, less accurate patrol
measurements or visual observations. They were taken into account as well,
at least at the first step of our analysis, but with different weights. We used four different
weights: 1: visually observed minima; 2: plate minima; 5: photographic normal
minima; and 10: photoelectric observations (both photomultiplier tube and CCD).
{\bf   \begin{table}
      \caption[]{Main parameters of the investigated stars. The
spectral types are taken from SIMBAD
Database\footnote{http://simbad.u-strasbg.fr/sim-fid.pl}.}
         \label{table:Starmainparams}
      \[
         \begin{array}{llllllllllll}
            \noalign{\smallskip}
            \hline
            \hline
            \noalign{\smallskip}
	    \mathrm{Name} & \mathrm{Sp} &P & \mathrm{T} & M_1 & M_2 & R_1 & R_2 & T_1 & T_2 & A &\mathrm{refs} \\
	    \noalign{\smallskip}
            \hline
            \noalign{\smallskip}
             & &\mathrm{d}& &\mathrm{M}_\odot&\mathrm{M}_\odot&\mathrm{R}_\odot&\mathrm{R}_\odot&\mathrm{K}&
               \mathrm{K}&\mathrm{R}_\odot&\\
            \noalign{\smallskip}
            \hline
            \noalign{\smallskip}
	    \mathrm{AB~And} & G5V^{a} & 0.33 & \mathrm{W} & 0.60 & 1.04 & 0.78 & 1.03 & 5450 & 5798 & 2.37 & 1 \\
	    \mathrm{OO~Aql} & G5V & 0.51 & \mathrm{A} & 1.04 & 0.88 & 1.39 & 1.29 & 5700 & 5560 & 3.33 & 2\\
	    \mathrm{DK~Cyg} & A6V &0.47 & \mathrm{A} & 1.74 & 0.53 & 1.71 & 0.99 & 7351 & 7200 & 3.34 & 1\\
	    \mathrm{V566~Oph}&F4V &0.41 & \mathrm{A} & 1.56 & 0.41 & 1.51 & 0.86 & 6700 & 6618 & 2.91& 3\\
	    \mathrm{U~Peg}  & G2V &0.37 & \mathrm{W} & 1.15 & 0.38 & 1.22 & 0.74 & 5860 & 5785 & 2.52& 4\\
	    \noalign{\smallskip}
            \hline
            \noalign{\smallskip}
          \end{array}
	  \]
	  \small{1: \citet{baranetal04}; 2: \citet{hrivnak89}; 3: \citet{niarchosetal93};
	   4: \citet{pribullavanko02} \\
           a: More recently \citet{pychetal04} found a mean spectral type of G8V}
\end{table}}
\subsection{\object{AB Andromedae} \label{ABAnd}}
This overcontact binary that belongs to the W subclass has been very well observed
since its discovery almost 80 years ago. A comprehensive analysis of the
photometric data obtained between 1968 and 1995 was recently
published by \citet{djurasevicetal00}, who concluded that the light-curve
variations can be explained well by variation of the spot activity. Two detailed
period investigations were carried out for this system in the last decade. Although
two different methods were applied in those papers. Both \citet{kalimerisetal94} and
\citet{borkohege96} detected the combination of secular period variation and cyclic
period change with a period which is close to the length of the observational interval.
This fact makes it especially interesting to carry out a new O--C study with
an extended data series which is longer by about 10 years.

\begin{figure}
   \centering
   \resizebox{\hsize}{!}{\includegraphics{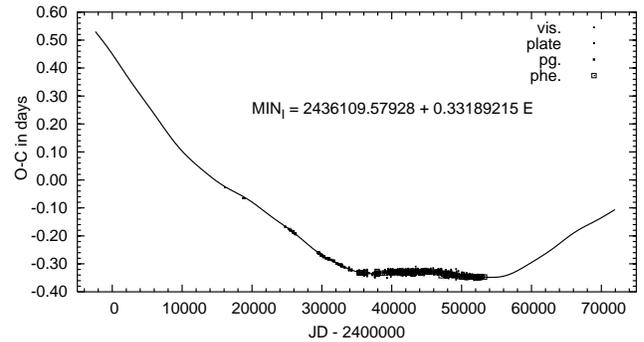}}
   \caption{The O--C curve of the eclipsing binary \object{AB And}, with the linear ephemeris taken
from the GCVS \citep{gcvs98} (above), and the simultaneous fit with two
fundamental frequencies. We plotted the visual points, too, although these were
discarded from the fitting procedure (see text for details).}
\label{Fig:O-CABAndossz}
    \end{figure}

\begin{figure}
   \centering
   \resizebox{\hsize}{!}{\includegraphics{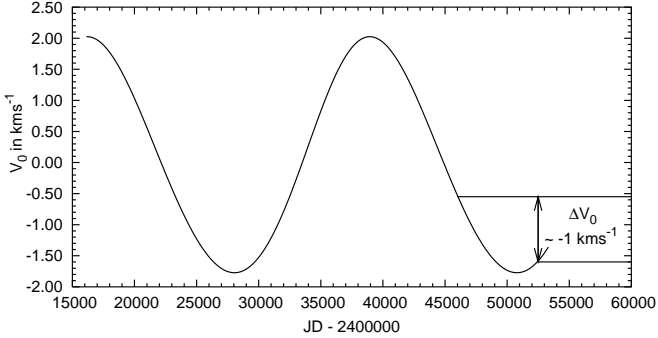}}
   \caption{The calculated systemic radial velocity ($V_0$) variation of \object{AB And} according to
   the LITE solution listed in Table~\ref{table:ABAnd}. We marked the positions of the radial velocity measurements
   of \citet{hrivnak88} and \citet{pychetal04}. See text for details.}
\label{Fig:RadABAnd}
    \end{figure}

  The analysed O--C data now cover 37\,181 days from HJD 2\,416\,103 to HJD 2\,453\,284.
As the photographic and photoelectric data cover all the interval well, apart from
the approx. 6\,000 day-long gap after the first three data, we omitted the visual data
in our analysis.
A weighted LSQ fit of Eq.~\ref{eq:O-Cgeneral}, with one fundamental frequency and with its
first harmonic, resulted in a very reasonable fit. Nevertheless, after subtraction of
this solution we found a second, small-amplitude quasi-periodic modulation with a
period of $P_\mathrm{mod}\approx7\,000$ days. In order to investigate this possible
modulation, we repeated our LSQ fitting procedure with two fundamental frequencies.
At the smaller frequency (i.e. longer period) we also took the first harmonic into account, 
while at the higher frequency we did not fit any harmonics.
Our final solution can be seen in Fig.~\ref{Fig:O-CABAndossz}.
Supposing that the longer period periodic term arises
from a light-time orbit, the orbital elements of the wide orbit of the binary around the
centre of mass of the hierarchical triple system can be calculated easily. The results
are listed in Table~\ref{table:ABAnd}.
   \begin{table}
      \caption[]{LITE solution for \object{AB And}}
         \label{table:ABAnd}
      \[
         \begin{array}{lll}
            \noalign{\smallskip}
            \hline
            \hline
            \noalign{\smallskip}
            \multicolumn{3}{l}{\mathrm{MIN}_\mathrm{I}=2436109.265216(2)+0.3318897(25)E+3.99(1)\times10^{-11}E^2}\\
            \noalign{\smallskip}
            \hline
            \noalign{\smallskip}
            P' & (\mathrm{day}) & 22\,742 \\
            e' &                & 0.10(1) \\
            \omega'& (\degr)    & 315(7) \\
            \tau' & \mathrm{(HJD)} & 2\,436\,358(448) \\
            a_{12}\sin{i'} & \mathrm{AU} & 3.93(4) \\
            f(m_3) & \mathrm{M}_\odot & 0.0158(5) \\
            K_{12} & \mathrm{kms}^{-1} & 1.90 \\
            \noalign{\smallskip}
            \hline
            \noalign{\smallskip}
            m_3 (\mathrm{M}_\odot)&i'=90\degr&0.40\\
            &i'=60\degr&0.48 \\
            &i'=30\degr&0.95\\
            \noalign{\smallskip}
            \hline
         \end{array}
      \]
   \end{table}
   \begin{figure}
   \centering
   \resizebox{\hsize}{!}{\includegraphics{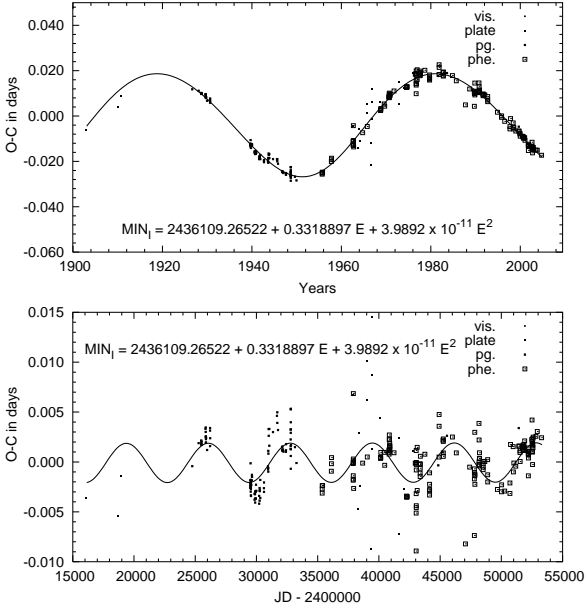}}
   \caption{{\bf Top: The} LITE fit with the period $P'=22\,742$ days. In order to see it
better we removed the parabolic term, as well as the shorter period sinusoidal component.
Bottom: The short term ($P_\mathrm{mod}=6\,695$ days) modulation on the O--C curve.
For the better seeing we removed the other components of our final solution.}
\label{Fig:O-CABAndkomp}
    \end{figure}
Nevertheless, the calculated minimal mass of the hypothetical tertiary
suggests that, if the tertiary is not a degenerate object, it has to produce a significant
amount of third light, which should be observed. In contradiction to this statement,
none of the previous light-curve solutions (at least those known for us)
contains any third light. On the other hand, we have to keep in mind that,
due to the complex effect of the individual fitting parameters on the light curve
solutions of contact eclipsing binaries, one can obtain similarly satisfactory approximation
of the same problem both with and without assuming third light; thus it would be necessary
to reanalyse the light curves of \object{AB And} with third light. Furthermore, slight indirect evidence for the motion of the binary in
a triple system arises from the comparison of the systemic velocity ($V_0$) of the system in the
latest two radial velocity surveys. While \citet{hrivnak88} obtained $V_0=-24.6\pm0.9$ km s$^{-1}$ 
from the JD 2\,445\,886--2\,446285 measurements, the most recent study of \citet{pychetal04} around JD 2\,452\,500
resulted in $V_0=-27.5\pm0.7$ km s$^{-1}$. In Fig.~\ref{Fig:RadABAnd} we plotted the calculated systemic radial velocity
variation of the system due to the revolution around the centre of mass of the triple system. As can be seen,
our solution suggests a $\Delta V_0\approx-1$ km s$^{-1}$ variation between the above-mentioned two sets of observations,
which is not all that far from the measured difference.
Another possibility is to explain the cyclic variation by magnetic
activity (see Sect.~\ref{Subsec:magcyc}).
\subsection{\object{OO Aquilae} \label{OOAql}}
This high mass-ratio A-type overcontact binary was discovered by \citet{hoffleit32}. It belongs to the
small group of longer period contact binaries which are supposed to have evolved into the contact phase in the
(astronomically) near past \citep[see e.~g.][and references therein]{hrivnaketal01}. Spectroscopic study
by \citet{hrivnaketal01} revealed strong Mg II h \& k emission lines varying with time, as clear evidence
of chromospheric activity, although they also pointed out that \object{OO~Aql} is a less active system than other W UMa-type ones.
The most detailed period study in the recent literature was published by \citet{demircangurol96}. They concluded that the most
probable two scenarios for the large structure period variation were an abrupt period jump with
$\Delta P\approx-0.6$s around 1963 or a sinusoidal variation with a period of $P'\approx90$ yr, although
this second statement was quite ambiguous, as the total curve covered only one maximum without any minima.
They also mentioned the possibility of a parabolic representation. Nevertheless, in the present evolutionary
state of the system, it could be expected that the binary evolves toward a smaller mass-ratio state, which should
have resulted in an increasing period as was clearly in contradiction with the 1995 state.
\citet{demircangurol96} also investigated the presence of shorter, small amplitude fluctuations and
concluded that these variations did not show simple periodic nature.

\begin{figure}
   \centering
   \resizebox{\hsize}{!}{\includegraphics{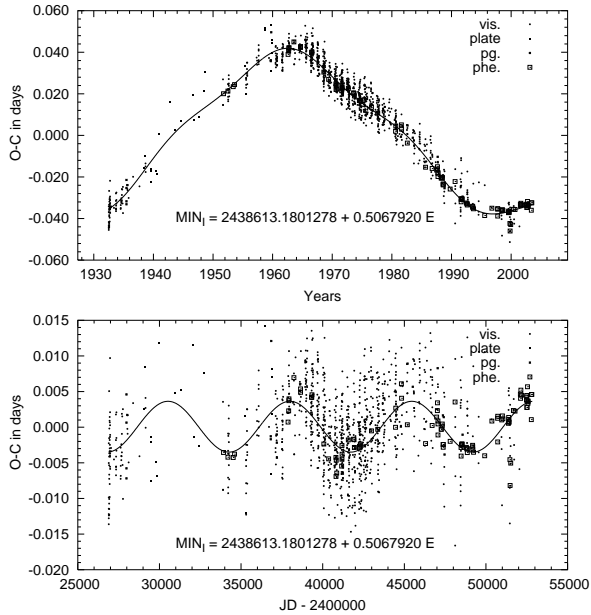}}
   \caption{The O--C curve of the eclipsing binary \object{OO Aql} with the calculated linear ephemeris and
   the simultaneous fit with two fundamental frequencies (above), and the smaller period fluctuations after
   the subtraction of the longer period curve. The Fourier-fits were carried out simultaneously.}
\label{Fig:O-COOAql}
    \end{figure}

Since this aforementioned study, the picture has changed significantly. Around 1996 the period of the system
also varied (see top panel in Fig.~\ref{Fig:O-COOAql}). Our O--C data extend on a 25\,919 day-long interval between
HJD 2\,426\,892 and 2\,452\,811. Although it cannot be excluded that this is a second
abrupt period change, we may also suspect that the large amplitude sinusoidal variation
reached its minimum. Consequently, we applied our fit with
two fundamental periods: one (with its first harmonic) for the long period sinusoidal variation,
and one for the short period fluctuations, which clearly reveals periodic behaviour in contrast to 
the statement of the previous authors. The simultaneous fit is plotted in the top panel of Fig.~\ref{Fig:O-COOAql},
while in the bottom panel we illustrate the periodic nature of O--C lacking long period terms.
Assuming that the $P'\approx75$yr variation arises from the presence of a more distant tertiary component, the
orbital elements of the binary in the triple system are listed in Table~\ref{table:OOAql}.
   \begin{table}
      \caption[]{LITE solution for \object{OO Aql}}
         \label{table:OOAql}
      \[
         \begin{array}{lll}
            \noalign{\smallskip}
            \hline
            \hline
            \noalign{\smallskip}
            \multicolumn{3}{l}{\mathrm{MIN}_\mathrm{I}=24438613.18013(1)+0.5067920(14)E}\\
            \noalign{\smallskip}
            \hline
            \noalign{\smallskip}
            P' & (\mathrm{day}) & 27\,273 \\
            e' &                & 0.06(1) \\
            \omega'& (\degr)    & 23(13) \\
            \tau' & \mathrm{(HJD)} & 2\,432\,894(970) \\
            a_{12}\sin{i'} & \mathrm{AU} & 6.56(5) \\
            f(m_3) & \mathrm{M}_\odot & 0.05078(11) \\
            K_{12} & \mathrm{kms}^{-1} & 2.62 \\
            \noalign{\smallskip}
            \hline
	    \noalign{\smallskip}
            m_3 (\mathrm{M}_\odot)&i'=90\degr& 0.70\\
            &i'=60\degr& 0.84\\
            &i'=30\degr& 1.77\\
            \noalign{\smallskip}
            \hline
         \end{array}
      \]
   \end{table}
As seen there, the minimal mass of the tertiary is again so large that (if not a degenerate object)
it should be seen via the light curve with significant third light contribution, as well as via its spectral lines.
Nevertheless, the previous light-curve studies give clear indirect evidence against
the presence of any significant third light; namely, that the inclination of the system without the supposition
of any third light was found to be close to $i=90\degr$ \citep{hrivnak89,djurasevicerkapic99}. As the third light
reduces the depth of the light minima, a given depth could be reproduced in its presence only by a larger
inclination value than in the case of no third light, which would be physically unrealistic for the present star.
Consequently, even if the 75 year-long periodic amplitude is real, an alternate explanation is needed for it.

\subsection{\object{DK Cygni} \label{DKCyg}}
Although the discovery of this A-type overcontact binary was reported in
the same paper \citep{guthnickprager27} than as was the above-mentioned
\object{AB~And}, and their brightness is also similar, this system was
less popular with the observers. \citet{paparoetal85} found a
continuous period increase in the system, which was confirmed by
\citet{wolfetal00}.  No spot activity was observed in this system as
a survey of the literature showed.

\begin{figure}
   \centering
   \resizebox{\hsize}{!}{\includegraphics{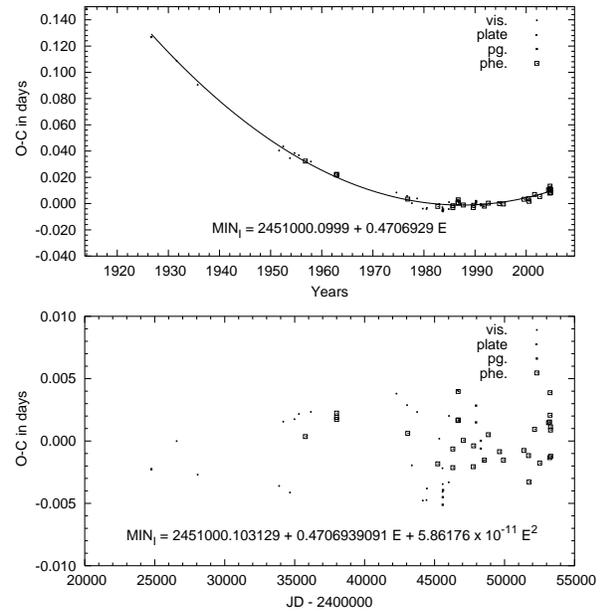}}
   \caption{The O--C curve of the eclipsing binary \object{DK Cyg}, with the linear ephemeris taken
from \citet{kissetal99} (above), and the residual curve after subtracting the parabola.}
\label{Fig:O-CDKCyg}
    \end{figure}

The collected times of minima data in our analysis cover 28\,542 days from HJD 2\,424\,760 to
HJD 2\,453\,302. In the analysis we also used seven recently obtained unpublished
minima, listed in Table~\ref{table:DKCygnewmin}.
Omitting the visual data, a weighted LSQ fit resulted in the following parabolic ephemeris (see Fig.~\ref{Fig:O-CDKCyg}):
\begin{equation}
MIN_I=2\,451\,000.1031+0.470693909E+5.862\times10^{-11}E^2,
\label{eq:DKCygO-C}
\end{equation}
\begin{table}
      \caption[]{New times of minima of \object{DK Cyg} observed by R. Csorv\'asi with CCD
      cameras mounted on the 40-cm Cassegrain telescope at Szeged Observatory, and the
      90-cm Schmidt telescope at Piszk\'estet\H o Mountain Station of Konkoly Observatory.}
         \label{table:DKCygnewmin}
      \[
         \begin{array}{lll}
            \noalign{\smallskip}
            \hline
            \hline
            \noalign{\smallskip}
            \mathrm{HJD}&\mathrm{Error}&\mathrm{Type}\\
            \noalign{\smallskip}
            \hline
            \noalign{\smallskip}
            2\,453\,223.4286 & \pm0.0011 &\mathrm{s}\\
            2\,453\,228.3681 & \pm0.0008 &\mathrm{p}\\
            2\,453\,246.4950 & \pm0.0013 &\mathrm{s}\\
            2\,453\,247.4346 & \pm0.0007 &\mathrm{s}\\
            2\,453\,285.326  & \pm0.002  &\mathrm{p}\\
            2\,453\,286.2657 & \pm0.0007 &\mathrm{p}\\
            2\,453\,302.2672 & \pm0.001  &\mathrm{p}\\
            \noalign{\smallskip}
            \hline
         \end{array}
      \]
   \end{table}
which is in good agreement with the previous results. Despite this fact,
we think that it is not without benefit to reanalyse the O--C of
this system, as well as any other eclipsing binaries, every 5-10 years
by the use of the most recent data points. This is first because, in
several cases, it cannot be decided from observations with gaps whether an abrupt period jump 
has occurred in a given time or if we see some continuous variation.
However, a more or less continuous monitoring could soon discover if something
interesting is happening, and would lead the observers (nowadays sometimes skilled
amateurs with advanced equipment like CCD cameras) to observe the given binary more frequently.
Second, returning to this particular system, the constancy of the period variation
(at least in the last 80 years), and the lack of any other features from the O--C curve
needs some explanation, especially because in most cases the O--C curves of the contact binaries
show more complex behaviour.

\subsection{\object{V566 Ophiuchi} \label{V566Oph}} %
This A-type contact binary was discovered by \citet{hoffmeister35}.
Although it was one of the most popular contact binaries during its
``first 60 years'', it seems to have been neglected in the past decade.
V566 Oph did have a stable light curve and constant period in the
past \citep{bookmyer76, eaton86}. After 1983 the depths of eclipse were
increased \citep{lafta85}. The behaviour of V566 Oph is not understood,
so it requires further study. Due to the light curve changes, the
system seems to be a rather active star.

The latest period analysis was carried out by \citet{qian01b}, who fitted
a least-squares parabolic ephemeris and, after subtracting this parabola,
he also fitted a pure sinusoidal term, with an amplitude of $0\fd0037$ d
and a period of $20.4$ yr. He interpreted this latter component as the
trace of a possible third companion.

As already mentioned in Sect.~\ref{Sec:O-Cgen}, the non-simultaneous fitting of the secular
and periodic terms might give inaccurate results. This was the main reason we repeated this
earlier O--C analysis; the other reason, of course, is that some new minima times have been obtained since then.
Our times of minima data cover the interval HJD 2\,434\,179--2\,452\,854. As was mentioned earlier, there are
only very few recent minima. We only found two within the last 7 years. In the present analysis we omitted the visual minima, because first
the total observing interval is satisfactorily covered by photoelectric minima and, second, the number of the visual minima
is so small that their statistical significance, which seems to be evident e.~g. in the case of \object{AB And}
or \object{U Peg}, fails here.

\begin{figure}
   \centering
   \resizebox{\hsize}{!}{\includegraphics{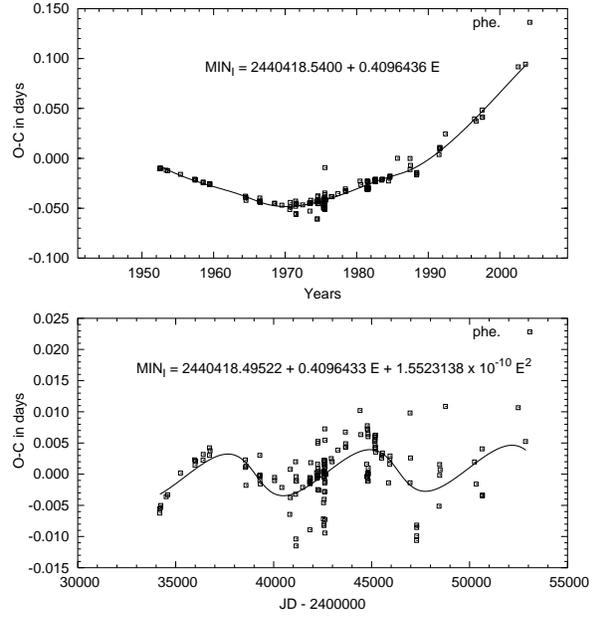}}
   \caption{The O--C curve of the eclipsing binary \object{V566 Oph}, with the linear ephemeris taken
from the GCVS \citep{gcvs98} (above), and the LITE fit on the parabola-subtracted curve.
The quadratic and Fourier-fits were carried out simultaneously.}
\label{Fig:O-CV566Oph}
    \end{figure}

   After omission of four evidently bad photoelectric data, we simultaneously fitted a quadratic polynomial
and the Fourier-terms up to the first harmonics. The best fit is plotted in Fig.~\ref{Fig:O-CV566Oph}, while the calculated LITE
orbit parameters are listed in Table~\ref{table:V566Oph}.
   \begin{table}
      \caption[]{LITE solution for \object{V566 Oph}}
         \label{table:V566Oph}
      \[
         \begin{array}{lll}
            \noalign{\smallskip}
            \hline
            \hline
            \noalign{\smallskip}
            \multicolumn{3}{l}{\mathrm{MIN}_\mathrm{I}=2440418.49522(7)+0.4096433(22)E+1.55(11)\times10^{-10}E^2}\\
            \noalign{\smallskip}
            \hline
            \noalign{\smallskip}
            P' & (\mathrm{day}) & 7\,250 \\
            e' &                & 0.31(8) \\
            \omega'& (\degr)    & 180(14) \\
            \tau' & (HJD) & 2\,453\,590(267) \\
            a_{12}\sin{i'} & \mathrm{AU} & 0.63(5) \\
            f(m_3) & \mathrm{M}_\odot & 0.00065(13) \\
            K_{12} & \mathrm{kms}^{-1} & 1.00 \\
            \noalign{\smallskip}
            \hline
	    \noalign{\smallskip}
            m_3 (\mathrm{M}_\odot)&i'=90\degr& 0.14\\
            &i'=60\degr& 0.17\\
            &i'=30\degr& 0.30\\
            \noalign{\smallskip}
            \hline
         \end{array}
      \]
   \end{table}
Comparing our results with those of \citet{qian01b}, both the parameters of the parabola and the periodic
terms are similar, for two reasons. First, the period of the quasi-sinusoidal term is
less than half of the length of the data series; consequently, the periodicity can be calculated almost
independently of the quadratic term parameters. Second, the amplitude of this term is small
with respect to the secular period change rate.
Although in Table~\ref{table:V566Oph} we present the LITE solution for this almost 20-year period modulation,
a possible magnetic origin is also investigated in Sect.~\ref{Subsec:magcyc}.

\subsection{\object{U Pegasi} \label{UPeg}}
This W subtype system is one of the oldest known W UMa type binaries, as was already discovered
110 years ago \citep{chandler895}. In a recent study, \citet{djurasevicetal01}
investigated the long term photometric variations of the system over the time interval of
1950 to 1989. They concluded that the complex variations of the light-curves during these
four decades can be explained by the variable spot activity of the cooler component. The
latest period study was carried out by \citet{pribullavanko02}, who analysed the features
of the O--C diagram between 1894 and 2001. These authors explained the century-long period
decrease by conservative mass transfer from the more massive to the less massive component.
Furthermore, they found a short period ($P_\mathrm{mod}=18.8\pm3.9$ years), low amplitude oscillation.
This feature had also been reported earlier by \citet{zhaietal84}.

\begin{figure}
   \centering
   \resizebox{\hsize}{!}{\includegraphics{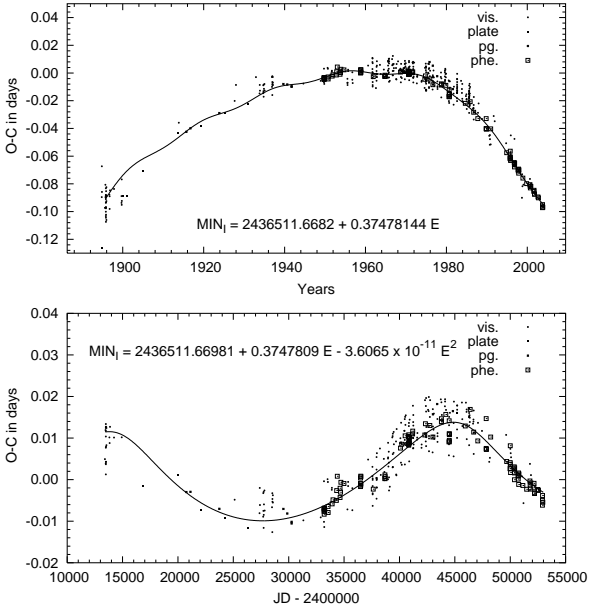}}
   \caption{The O--C curve of the eclipsing binary \object{U Peg} with our final solution, plotted against the
   linear ephemeris taken from the GCVS \citep{gcvs98} (above), and the pure LITE fit on the parabola
   subtracted curve. The quadratic and Fourier-fits were carried out simultaneously.}
\label{Fig:O-CUPeg}
    \end{figure}

\begin{figure}
   \centering
   \resizebox{\hsize}{!}{\includegraphics{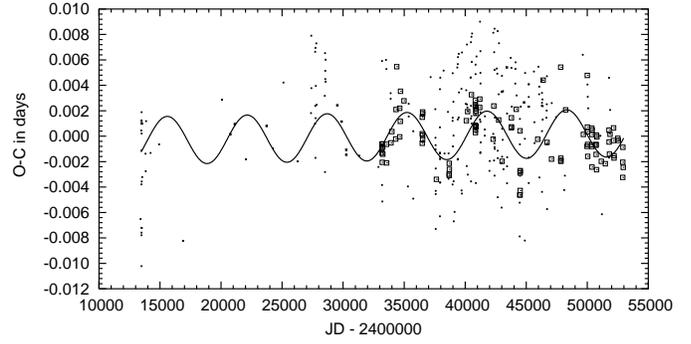}}
   \caption{The shorter period sinusoidal fit on the O--C of \object{U Peg}. For the sake of clarity,
   we subtracted the parabola, as well as the longer period LITE solution, from the original curve.}
\label{Fig:O-CUPegfeher}
    \end{figure}

Our O--C data cover 39\,856 days between HJD 2\,413\,094 (1894) and 2\,452\,950 (2003). These data are
plotted in the upper panel of Fig.~\ref{Fig:O-CUPeg}. We carried out our analysis in a similar way as
the one described in the case of \object{AB~And}; i.~e.
we fitted simultaneously a quadratic polynomial and two fundamental frequencies allowing the first harmonic at the longer
period as well. Our best fit gave the following two periods: $P_1\approx31\,059$ days,
$P_2\approx6\,544$ days. If the first period belongs to a light-time orbit due to the existence of an additional star
in a hierarchical triple (multiple) system the corresponding orbital elements and derived quantities can be
read in Table~\ref{table:UPeg}. The latter period is approximately the same as that
reported in the previous papers; what perhaps is more interesting, this is also
very similar to the length of the modulations in the previous stars. Nevertheless, in the present
case it is questionable that this is a real period variation.
As can be seen well in Fig.~\ref{Fig:O-CUPegfeher}, the total amplitude of this ``variation''
is only $A_2\approx0\fd003$, i.~e. smaller than the scatter of the photoelectric minima,
which is usual in W UMa binaries, as a consequence of the varying light-curve shapes distorted
by spots (see e.~g. \citealp{kalimerisetal02}). In our opinion it might be imagined that the presence of such a periodicity,
especially between 1960 and 1980, is only a result of the small number of photoelectric measurements.
However, the fact that similar periodicity was found in the other cases, too, makes it reasonable that
this feature might be real.
   \begin{table}
      \caption[]{LITE solution for \object{U Peg}}
         \label{table:UPeg}
      \[
         \begin{array}{lll}
            \noalign{\smallskip}
            \hline
            \hline
            \noalign{\smallskip}
            \multicolumn{3}{l}{\mathrm{MIN}_\mathrm{I}=2436511.6698(5)+0.3747809(5)E-3.61(10)\times10^{-11}E^2}\\
            \noalign{\smallskip}
            \hline
            \noalign{\smallskip}
            P' & (\mathrm{day}) & 31\,059 \\
            e' &                & 0.38(5) \\
            \omega'& (\degr)    & 112(9) \\
            \tau' & (HJD) & 2\,445\,935(569) \\
            a_{g12}\sin{i'} & \mathrm{AU} & 1.96(13) \\
            f(m_3) & \mathrm{M}_\odot & 0.00105(20) \\
            K_{12} & \mathrm{kms}^{-1} & 0.74 \\
            \noalign{\smallskip}
            \hline
            \noalign{\smallskip}
            m_3 (\mathrm{M}_\odot)&i'=90\degr&0.14 \\
            &i'=60\degr&0.17 \\
            &i'=30\degr&0.30\\
            \noalign{\smallskip}
            \hline
         \end{array}
      \]
   \end{table}
\section{Discussion}
The five investigated systems are distributed in the range of effective temperature from $T_\mathrm{eff}=7400$ down to $T_\mathrm{eff}=5400$,
which is the transition region where a convection zone develops in the envelope of the stars
toward the lower temperatures. Analysis of the O--C diagrams of the systems has revealed
different periodic variations in four cases besides the secular development of the orbital period.

\subsection{Secular period changes}\label{Subsec:seccyc}
Four of the five O--C curves show evidence of continuous orbital period change with a constant rate during the total observing interval.
The only exception was \object{OO~Aql}, although it is possible that in this
system only one (or two) abrupt period jump(s) obscures this behaviour.

The most usual explanation of this kind of secular period variation is mass exchange in the system.
The widely used approximating formula for the calculation of the mass exchange rate is as follows:
\begin{equation}
\dot{m}=-m_{12}\frac{q}{1-q^2}\frac{\dot{P}}{3P}\approx-m_{12}\frac{q}{1-q^2}\frac{2}{3}\frac{c_2}{c_1^2},
\label{eq:mass-exchange}
\end{equation}
where $c_1$, $c_2$ directly come from the O--C ephemerides in the form of Eq.~(\ref{eq:O-Cgeneral}).
Note that the equation requires constancy of the total mass, as well as of the total angular momentum.
The rate of the secular period change, as well as the calculated mass exchange rate, are tabulated in
Table~\ref{table:Starderivparams}.
   \begin{table}
      \caption[]{Derived parameters from the period variations}
         \label{table:Starderivparams}
      \[
         \begin{array}{llllllll}
            \noalign{\smallskip}
            \hline
            \hline
            \noalign{\smallskip}
	    \mathrm{Name} & \dot{P}_\mathrm{sec}/P & |\dot{m}| & P_\mathrm{mod} & A_\mathrm{mod} &  &  &  \\
	    \noalign{\smallskip}
            \hline
            \noalign{\smallskip}
	    & \times10^{-7}\mathrm{~(y}^{-1}\mathrm{)} & \times10^{-7}\mathrm{~(M}_\odot\mathrm{~y}^{-1}\mathrm{)} & \mathrm{(d)} & \mathrm{(d)} &  &  &  \\
	    \noalign{\smallskip}
            \hline
            \noalign{\smallskip}
	    \mathrm{AB~And} & 2.65 & 1.25 & 6\,695 & 0.0020 &  &  &  \\
	    \mathrm{OO~Aql} & -  & - & 7\,467 & 0.0036 &  &  & \\
	    \mathrm{DK~Cyg} & 1.93 & 0.49 & - & - &  &  &  \\
	    \mathrm{V566~Oph}&6.76 & 1.25 & 7\,250 & 0.0035 &  &  &  \\
	    \mathrm{U~Peg}  &-1.88 & 0.36 & 6\,544 & 0.0016 &  &  &  \\
	    \noalign{\smallskip}
            \hline
            \noalign{\smallskip}
          \end{array}
	  \]
\end{table}

Nevertheless, it is necessary to note, that in the case of \object{AB~And} and
\object{U~Peg}, the coefficients of the quadratic terms and the period and amplitude
of the longer period quasi-sinusoidal period variations are not independent of each other
as we will see later.

\subsection{Longer period cyclic variations}\label{Subsec:longpercyc}

In three cases the simultaneous fitting gave longer, as well as shorter, period cycles.
These longer scale variations
are roughly 62, 75, and 85 years long and naturally could be identified as a LITE, i.~e. effect
due to the presence of a third body. The corresponding LITE solution parameters were given
in the corresponding tables along with the detailed discussion of the systems.
The identification of these periods as LITE
solution can be considered only as a simple designation, and their real origin should still require a
careful analysis of all cases.

It should also be emphasised that these periods are in the same order as the total interval of the observations, and the
separation of the different effects in the analysis are doubtful. For example,
let us consider e.~g. the case of \object{AB~And}. In this case the fundamental Fourier-terms
are as follows:
\begin{equation}
f=0.015\sin(9.17\times10^{-5}E)-0.017\cos(9.17\times10^{-5}E)+...,
\label{Eq:fundtermsFour}
\end{equation}
which can be written as
\begin{equation}
f=-0.017+1.38\times10^{-6}E+7.15\times10^{-11}E^2+...,
\label{Eq:fundTay}
\end{equation}
where as seen well, the quadratic term is in the same order of magnitude than the
coefficient of the parabolic one. This fact is critical mainly in these two aboveümentioned cases
(\object{AB~Cas} and \object{U~Peg}),
as in both systems only the last half quasi-sines are covered well by good-quality photoelectric
data. On the other hand, despite this fact, we can stress that at least the order of magnitude of
the calculated secular period change is believed to be correct.
\subsection{Shorter period cyclic variations - manifestation of magnetic activity cycles? \label{Subsec:magcyc}}
It is an interesting fact that four of our five systems
show small amplitude cyclic fluctuations with very similar periods
(see Cols.~4-5 in Table~\ref{table:Starderivparams}).
The only exception is \object{DK Cyg}, the hottest in the sample, with A8 spectral-type components,
whilst the other four are later than F0 down to G5.

These modulation periods range from 6\,500 to 7\,500 days (18--20 years), which are quite similar to 
those observed in other types of stars showing magnetic activity. 
Comparing them to the longer period components of the detected variations,
where the observations cover only partly a cycle, these shorter cycles are well-covered
(several times, 2--4 cycles), as can be seen from the discussion of the individual systems.
Hence, the existence and characteristics of these cyclical variations in the orbital period are well
proved independently from any doubts about the other -- longer period -- components.

The startling resemblance of orbital period variations in these more or less similar systems
suggests some parallel in their origins and makes some common physical explanations likely.
In our opinion this feature might be indirect evidence of similar magnetic cycles
in the investigated binaries, and the detected variations might be caused by it.

We mainly refer to \citet{lanzarodono99}, who found a relation between the orbital period of a
binary system and the magnetic activity cycle by supposing synchronisation between the orbital
motion and rotation of the member stars:
\begin{equation}
\log P_\mathrm{mod}~[\mathrm{yr}] = 0.018 - 0.36(\pm0.10) \log \frac{2\pi}{P_\mathrm{orb}} [\mathrm{sec}].
\label{eq:PmodPorb}
\end{equation}
For the binaries studied here, this formula predicts $P_{\mathrm{mod}}\approx7\,900$,
9\,200, 8\,500, and 8\,200  days, respectively (note values very close to the length of the Sun's magnetic cycle).
These values are on the same order of magnitude as the observed ones. This supports
our conjecture that magnetic activity cycles could explain the observed short term orbital
period variations.
Furthermore, beyond our sample, \citet{awadallaetal04}
reported a similar periodicity of $P_\mathrm{mod}\approx6\,500$ days in the O--C curve of the
late-type (F8) W~UMa binary \object{AK~Her}, giving additional support to the explanation above.

The rate of this period variation can be expressed easily \citep[see e.~g.][]{applegate92} as
\begin{equation}
\frac{\Delta P}{P} = 2 \pi \frac{A_\mathrm{O-C}}{P_{\mathrm{mod}}}
\end{equation}
where ${\Delta P}/{P}$ is the relative period variation, $A_\mathrm{O-C}$ the
half-amplitude of the O--C variation, and $P_{\mathrm{mod}}$ the cycle length of
the magnetic activity.

In the case of \object{AB~And}, applying $P_{\mathrm{mod}} = 6\,700$
days and $A=0.002$
days, respectively, we get ${\Delta P}/{P} = 1.9\times10^{-6}\mathrm{d/d}$. The
variation of the gravitational quadrupole momentum is \citep{applegate92}:
\begin{equation}
\Delta Q = -\frac{1}{9} MR^2 \left(R/a\right)^{-2} \frac{\Delta P}{P}
\end{equation}
where $M$ is the mass of the active star, $R$ the radius of the active star,
and $a=2A$ the separation between the components. $A$ is the semi-major axis,
and note that in contact binaries the orbit is circular. A calculation with the
parameters of the primary and the secondary (see Table~\ref{table:Starmainparams}) gives
$\Delta Q_1 = 4.02 \times 10^{42}\mathrm{kgm}^2$
and $\Delta Q_2 = 6.97 \times 10^{42}\mathrm{kgm}^2$, respectively.
These values correspond to the typical values in active binary stars \citep{lanzarodono99}.
Due to the similar absolute dimensions of the other investigated binaries, the same calculations
give physically realistic results for them as well.

Nevertheless, we also have to emphasise that in a recent study \citet{qian03}
did not find clear evidence of decades time-scale period variations in similar systems.
He investigated the period variation of five contact binaries (\object{GW~Cep} (G3),
\object{VY~Cet} (G5V), \object{V700~Cyg} (G2V), \object{EM~Lac} (G8V), \object{AW~Vir} (G0V)) and
found a secular period change in every case, and he discovered that only two
of them show cyclic period variations (\object{VY~Cet} ($P_3 = 7.3$ years),
\object{V700~Cyg} ($P_3 = 39.8$ years)). It is an interesting fact that only 40
percent of these G spectral type systems have cyclic period variations,
despite expecting such a phenomenon in all cases.  Although both
samples (that of \citealt{qian03} and our) are small, and therefore not
representative, the controversion should be studied. Our own opinion is as
follows. The O--C diagrams of the three remaining systems (\object{GW~Cep}, \object{EM~Lac}
and \object{AW~Vir}) consist in only a few points, and there are large gaps between
the data. This is due to the fact that these systems were underobserved in
the past; therefore, the present observational material is enough to
detect the secular period variation only -- in spite of applying all
available minima. Gaps and scattered observations (like visual and
photographic ones) obstruct the demonstration of such cyclic variations.

As mentioned already, the spectral type of the systems is later than F0, where we have found
short period variations in the O--C diagrams, the earlier type \object{DK~Cyg} does not show this
feature. This type of behaviour is pretty similar to the observed one among the Algol-type RS~CVn
systems \citep[see][]{hall98} so that the magnetic activity can be observed only in systems later
than F5. Because the magnetic dynamo theories are based on a strong connection with the presence
of a convection zone in the outer envelope predicting magnetic activity for them. Since
the convection zones in the envelope appear in lower mass MS and cooler stars, this type
of relation is a trivial requirement but, of course, the strict border lines can be different.
Although our sample is not a systematic one, it indicates the existence
of a separation border somewhere around the spectral type F0.

\section{Conclusions}

The orbital period is very sensitive to any effect and it could be measured
with very high precision down to $10^{-9}$ part of the period, and we have
a lot of series of minimum observations, in some cases back to the 19th century.
In this study we have investigated the O--C diagrams of five W~UMa systems from A8 to G5
spectral type components. In four cases a short time scale ($\simeq 20$ years) cyclic variation
of the orbital period was found. All these systems contain later than F0 spectral-type
components, which means that they have deep convection zones. As we
showed the very similar modulation periods that were observed seem to be independent
of any LITE caused by a third body, but a consistent explanation of their cause might also be
a magnetic activity cycle.

The mechanism proposed by \citet{applegate92} seems to be responsible for the observed
variations, and the investigation of close binary systems from this aspect could
be an important tool for studying the magnetic properties of stars in general. However, contact
binary stars are generally faint and the geometry very complex, so that it is very difficult to apply
these methods until further further developments \citep[see e.~g.][]{hendrymochnacki00}.

Of course, these studies have to be complemented by detailed analyses of the systems from other
viewpoints, e.~g. observations of the time variable features of their spectra, modelling of
time development of their light curves, etc. For example, to check our hypothesis about the
magnetic activity, we should follow in time the spotted light curve models of the systems
determining their spottedness--index and its correlation with the observed O--C variations.
These works are in progress and will be published in due time.

\begin{acknowledgements}  This research made use of NASA's Astrophysics
Data System Abstract and Article Service. This work was partly
supported by the Hungarian OTKA Grants T~034551 and T~042509. We thank Dr L. Szabados for his kind
help and advice during the preparation of the manuscript.
\end{acknowledgements}


\vskip1cm

\bibliographystyle{aa}

\begin{thebibliography}{}


\bibitem[Applegate(1992)]{applegate92} Applegate, J.H., 1992, \apj, 385, 621

\bibitem[Awadalla et al.(2004)]{awadallaetal04} Awadalla, N., Chochol, D., Hanna, M., \& Pribulla, T.,
   2004, Contrib. Astron. Obs. Skalnat\'e Pleso, 34, 20

\bibitem[Baran et al.(2004)]{baranetal04} Baran, A, Zola, S, Rucinski, S.M. et al., 2004, AcA, 54, 195

\bibitem[Barnes et al.(2004)]{barnesetal04} Barnes, J.R., Lister, T.A., Hilditch, R.W., \& Collier Cameron, A.,
   2004, \mnras, 348, 1321

\bibitem[Bookmyer(1976)]{bookmyer76} Bookmyer, B. B.,
   1976, \pasp, 88, 473

\bibitem[Borkovits \& Heged\"us(1996)]{borkohege96} Borkovits, T., \& Heged\"us, T.,
   1996, \aaps, 120, 63

\bibitem[Borkovits et al.(2002)]{borkoetal02} Borkovits, T., Csizmadia, Sz., Heged\"us, T. et al.,
   2002, \aap, 392, 895

\bibitem[Borkovits et al.(2003)]{borkoetal03} Borkovits, T., \'Erdi, B., Forg\'acs-Dajka, E., \& Kov\'acs, T.,
   2003, \aap, 398, 1091

\bibitem[Chandler(1895)]{chandler895} Chandler, S.C., 1895, \aj, 15, 185

\bibitem[Csizmadia et al.(2004)]{csizmadiaetal04} Csizmadia, Sz., Patk\'os, L., Mo\'or, A., K\"onyves, V.,
   2004, \aap, 417, 745

\bibitem[Demircan et al.(1991)]{demircanetal91} Demircan, O., Derman, E., \& Akalin, A., 1991,
   \aj, 101, 201

\bibitem[Demircan \& G\"urol(1996)]{demircangurol96} Demircan, O., \& G\"urol, B., 1996,
   \aaps, 115, 333

\bibitem[Djura\v sevi\'c \& Erkapi\'c(1999)]{djurasevicerkapic99} Djura\v sevi\'c, G., \& Erkapi\'c, S.,
   1999, \apss, 262, 305

\bibitem[Djura\v sevi\'c et al.(2000)]{djurasevicetal00} Djura\v sevi\'c, G., Rovithis-Livaniou, H., \&
   Rovithis, P., 2000, \aap, 364, 543

\bibitem[Djura\v sevi\'c et al.(2001)]{djurasevicetal01} Djura\v sevi\'c, G., Rovithis-Livaniou, H.,
   Rovithis, P., Erkapi\'c, S., \& Milovanovi\'c, N., 2001, \aap, 367, 840

\bibitem[Eaton(2001)]{eaton86} Eaton, J. A.,
   1986, AcA, 36, 275

\bibitem[Guthnick \& Prager(1927)]{guthnickprager27} Guthnick, P., \& Prager, R., 1927,
   AN, 229, 455

\bibitem[Hall (1989)]{hall98} Hall, D.S., 1989, \ssr, 50, 219

\bibitem[Hazlehurst(2001)]{hazlehurst01} Hazlehurst, J., 2001, The Obs., 121, 86

\bibitem[Heintz(1993)]{heintz93} Heintz, W.D., 1993, \pasp, 105, 586

\bibitem[Hendry \& Mochnacki(1998)]{hendrymochnacki98} Hendry, P. D., \& Mochnacki, S. W., 1998,
   \apj, 504, 978

\bibitem[Hendry \& Mochnacki(2000)]{hendrymochnacki00} Hendry, P. D., \& Mochnacki, S. W., 2000,
   \apj, 531, 467

\bibitem[Hershey(1975)]{hershey75} Hershey, J.L., 1975, \aj, 80, 662

\bibitem[Hoffleit(1932)]{hoffleit32} Hoffleit, D., 1932, Harvard Obs. Bull, 807, 9

\bibitem[Hoffmeister(1935)]{hoffmeister35} Hoffmeister, C., 1935, AN, 255, 401

\bibitem[Hrivnak(1988)]{hrivnak88} Hrivnak, B.J., 1988, \apj, 335, 319

\bibitem[Hrivnak(1989)]{hrivnak89} Hrivnak, B.J., 1989, \apj, 340, 458

\bibitem[Hrivnak et al.(1995)]{hrivnaketal95} Hrivnak, B.J., Guinan, E.F., \& Lu,  W.,
    1995, \apj, 455, 300

\bibitem[Hrivnak et al.(2001)]{hrivnaketal01} Hrivnak, B.J., Guinan, E.F., DeWarf, L.E., \&
    Ribas, I., 2001, \aj, 121, 1084

\bibitem[Kalimeris et al.(1994)]{kalimerisetal94} Kalimeris, A., Rovithis-Livaniou, H.,
   Rovithis, P., et al., 1994, \aap, 291, 765

\bibitem[Kalimeris et al.(2002)]{kalimerisetal02} Kalimeris, A., Rovithis-Livaniou, H.,
   Rovithis, P., et al., 2002, \aap, 387, 969

\bibitem[Kasz\'as et al.(1998)]{kaszasetal98} Kasz\'as, G., Vink\'o, J., Szatm\'ary, K., et al.
   1998, \aap, 331, 231

\bibitem[K\"ahler(1986)]{kahler86} K\"ahler, H., 1986, MitAG, 67, 85

\bibitem[K\"ahler(2004)]{kahler04} K\"ahler, H., 2004, \aap, 414, 317

\bibitem[Kholopov et al.(1998)]{gcvs98} Kholopov, P.N., Samus, N. N., Frolov, M. S. et al., 1998,
   Combined General Catalogue of Variable Stars, 4.1

\bibitem[Kiss et al.(1999)]{kissetal99} Kiss, L.L., Kasz\'as, G., F\H ur\'esz, G., \& Vink\'o, J.,
   1999, IBVS 4681

\bibitem[Kopal(1978)]{kopal78} Kopal, Z., 1978, Dynamics of Close Binary Systems (D. Reidel, Dordrecht)

\bibitem[Kreiner(1977)]{kreiner77} Kreiner, J.M., 1977, The Interaction of Variable Stars with
   their Environment, Proceedings of IAU Colloq. 42, held in Bamberg, September 6-9, 1977, Bamberg:
   Remeis-Sternwarte, edited by Rudolf Kippenhahn, J. Rahe, and W. Strohmeier., p.393

\bibitem[Kreiner et al.(2003)]{kreineretal03} Kreiner, J. M., Rucinski, S. M., Zola, S. et al., 2003,
    \aap, 412, 465

\bibitem[Lafta \& Grainger(1985)]{lafta85} Lafta, S. J., \& Grainger, J.
F., 1985, \apss 114, 23

\bibitem[Lanza \& Rodon\'o(1999)]{lanzarodono99} Lanza, A. F., \& Rodon\'o, M., 1999, \aap, 349, 887

\bibitem[Lanza \& Rodon\'o(2002)]{lanzarodono02} Lanza, A. F., \& Rodon\'o, M., 2002, AN, 323, 424

\bibitem[Lubow \& Shu(1977)]{lubowshu77} Lubow, S. H., \& Shu, F. H., 1977, \apj, 216, 517

\bibitem[Lucy(1976)]{lucy76} Lucy, L.B., 1976, \apj, 205, 208

\bibitem[Maceroni \& van't Veer(1996)]{maceronivantveer96} Maceroni, C., van't Veer, F. 1996,
   \aap, 311, 523

\bibitem[Maceroni et al.(1994)]{maceronietal94} Maceroni, C., Vilhu, O., van't Veer, F., \& van Hamme, W.,
   1994, \aap, 288, 529

\bibitem[Mochnacki(1981)]{mochnacki81} Mochnacki, 1981, \apj, 245, 650

\bibitem[Niarchos et al.(1993)]{niarchosetal93} Niarchos, P.G., Rovithis-Livaniou, H., \& Rovithis, P.,
   1993, \apss, 203, 197

\bibitem[Papar\'o et al.(1985)]{paparoetal85} Papar\'o, M., Hamdy, M. A., \& Jankovics, I., 1985,
   IBVS No. 2838

\bibitem[Pribulla \& Va\v nko(2002)]{pribullavanko02} Pribulla, T., \& Va\v nko, M., 2002,
   Contrib. Astron. Obs. Skalnat\'e Pleso, 32, 79

\bibitem[Pych et al.(2004)]{pychetal04} Pych, W., Rucinski, S.M., DeBond, H., et al., 2004, \aj, 127, 1712

\bibitem[Qian(2001a)]{qian01a} Qian, S., 2001a, \mnras, 328, 635

\bibitem[Qian(2001b)]{qian01b} Qian, S., 2001b, \mnras, 328, 914

\bibitem[Qian(2003)]{qian03} Qian, S., 2003, \mnras, 342, 1260

\bibitem[Rovithis-Livaniou et al.(1999)]{rovithislivaniouetal99} Rovithis-Livaniou, H., Kranidiotis, A.,
   Fragoulopoulou, E., Sergis, N., \& Rovithis, P., 1999, IBVS, 4713

\bibitem[Stepien et al.(2001)]{stepienetal01} Stepien, K., Schmitt, J.H.M.M., \& Voges, W.,
   2001, \aap, 370, 157

\bibitem[van't Veer(1979)]{vantveer79} van't Veer, F., 1979, \aap, 80, 287

\bibitem[van't Veer \& Maceroni(1989)]{vantveermaceroni89} van't Veer, F., Maceroni, C.,
   1989, \aap, 220, 128

\bibitem[Webbink(1976)]{webbink76} Webbink, R. F., 1976, \apj, 209, 829

\bibitem[Wolf et al.(2000)]{wolfetal00} Wolf, M., Molik, P., Hornoch, K., \&
   Sarounov\'a, L., 2000, \aaps, 147, 243

\bibitem[Zhai et al.(1984)]{zhaietal84} Zhai, D.-S., Leung, K.-C., Zhang, R.-X., 1984,
   \aaps, 57, 487

\end{thebibliography}

\end{document}